\documentclass{article}
\usepackage{hiph-preprint}
\volnumber{22} \issuenumber{1} \edyear{2005}                             
\frompage{000} \topage{000}                                              
\recrevdate{1 January 2005}                                              
\def\rightmark{Dissipation of low-$Q^2$ partons}
\def\leftmark{R. Ray and M. Daugherity}

\def\bea{\begin{eqnarray}}
\def\eea{\end{eqnarray}}
\newcommand{\insertplotx}[1]{\begin{center}\leavevmode\epsfysize=4.8cm
\epsfbox{#1}\end{center}}
\newcommand{\insertplotxx}[1]{\begin{center}\leavevmode\epsfysize=3.3cm
\epsfbox{#1}\end{center}}
 
\title{Dissipation and fragmentation of low-$Q^2$ scattered partons in Au-Au 
collisions at RHIC} 
\authors{
{R. L. Ray and M. Daugherity, STAR Collaboration %
\index{Ray, R.} 
\index{Daugh, M.} 
}\\[2.812mm]
{\normalsize
\hspace*{-0pt}Department of Physics, The University of Texas at Austin,\\ 
Austin, Texas 78712, USA\\[0.2ex] 
}}
 
\abstract{
Two-particle correlations and event-wise fluctuations in transverse momentum $p_t$ are reported for Au-Au
collisions at $\sqrt{s_{NN}}$ = 62 and 200 GeV on pseudorapidity $\eta$
and azimuth $\phi$.
Distributions of all pairs of particles (no leading trigger particle)
reveal jet-like correlations, or peaks at pair-wise opening angles of
order 1 radian or less.
The width of this {\em same-side} correlation peak
increases dramatically on pseudorapidity and decreases on
azimuth for increasing collision centrality.
Evolution of the same-side peak with centrality suggests
dissipation of low-$Q^2$ partons via strong coupling to an expanding bulk
medium.  $p_t$ correlations, which provide access to temperature
and/or velocity distributions in the colliding system, are also presented.}
\keyword{correlations, fluctuations, Au-Au collisions, minijets}

\PACS{24.60.Ky, 25.75.Gz}
 
\makeindex
\begin{document}
 
\maketitle

\section{Introduction}\label{intro}
Theoretical descriptions of relativistic heavy-ion collisions predict
abundant low-$Q^2$ gluon production in the early stages of the collision
with rapid parton thermalization driving the formation of a colored
medium~\cite{theor0,theor1,theor2}. Remnants of these semi-hard processes
may survive to the final, decoupling stage owing to the finite size and
lifetime of the collision volume.  If so, there should be contributions
to nonstatistical event-wise fluctuations of mean-$p_t$ $\langle p_t \rangle$
and two-particle correlations. Large
nonstatistical $\langle p_t \rangle$ fluctuations have been reported for
Au-Au collisions at 130 GeV~\cite{meanptprc} and 200 GeV~\cite{Phenixpt}
which are much larger than was observed at the SPS~\cite{ceres,anticic}.
Study of the related two-particle correlations facilitates interpretation
of event-wise fluctuations in terms of underlying dynamics.
In this paper we report preliminary two-particle correlations for Au-Au
collisions at 62 GeV on ($\eta,\phi$) and $\langle p_t \rangle$ fluctuations and $p_t$
correlations for Au-Au at 200 GeV~\cite{ptscale} from the STAR Collaboration.
Low-$Q^2$ partons are emphasized
in order to probe
nonperturbative QCD medium effects.
These data complement the two-particle correlations
for Au-Au collisions at 130 GeV reported in~\cite{mtxmtCI,axialCI}.

Data for the present analysis
were obtained with the STAR detector~\cite{star} where
charged particles were accepted in $|\eta| \leq 1.3$, $2\pi$ azimuth, and
$p_t \geq 0.15$~GeV/$c$.
Corrections were made for two-track inefficiencies and charge-sign dependent
cuts were applied to minimize quantum and Coulomb correlation contributions.
These cuts do not significantly affect the correlation
structures shown here. Cuts based on $dE/dx$ in the STAR Time Projection
Chamber tracking detector
were applied to reduce photon conversion $e^{\pm}$
contamination. See Ref.~\cite{axialCI} for discussion of the various cuts.

\section{Autocorrelations}\label{autocorr}
In conventional jet analysis event-wise concentrations of transverse momentum or
energy localized on angle variables $(\eta,\phi)$ are identified.
In heavy ion collisions, where such identification is impractical,
jet studies are based on a high-$p_t$ `leading particle' which may
estimate a parton momentum direction and some fraction of its magnitude.
However, for low-$Q^2$ partons identification of a `leading particle' is
not possible.  Furthermore the number of correlated hadrons from each 
low-$Q^2$ parton is expected to be relatively small while the number of
such partons
per event is large (of order tens).  Observation of these relatively soft
correlated hadrons 
is accessed via {\em autocorrelations} which are
well-known in signal processing disciplines to enable weak but repetitive
signals, which cannot be detected individually, to be accurately measured
in aggregate.
An autocorrelation is a projection
{\em by averaging}
from subspace $(x_1,x_2)$ onto difference variable $x_\Delta \equiv
x_1 - x_2$.
The autocorrelations reported in~\cite{mtxmtCI,axialCI}
and here do not require a leading- or trigger-particle, but instead use
all particle pairs within the acceptance.

The two-particle correlation density
per final state particle is defined by
\bea
\Delta\rho/\sqrt{\rho_{mix}} & \equiv &
[\rho_{sib}(\vec{p}_1,\vec{p}_2) - \rho_{mix}(\vec{p}_1,\vec{p}_2)]
/\sqrt{\rho_{mix}},
\label{RLR_Eq1}
\eea
where $\rho_{sib}(\vec{p}_1,\vec{p}_2)$ is
the object distribution comprised
of particle pairs from single events ({\it sibling} pairs)
and the reference distribution, $\rho_{mix}(\vec{p}_1,\vec{p}_2)$,
consists of pairs formed by sampling each
particle of the pair from two different but similar events
({\it mixed-event} pairs).
Studies of measured two-particle
correlation projections onto
subspaces ($p_{t1}$ vs $p_{t2}$),
($\eta_1$ vs $\eta_2$) and ($\phi_1$ vs $\phi_2$)
indicate that the principle dependences for Au-Au collision data are on
($p_{t1} + p_{t2}$), ($p_{t1} - p_{t2}$), and the
differences
$\eta_{\Delta} \equiv \eta_1 - \eta_2$ and
$\phi_{\Delta} \equiv \phi_1 - \phi_2$.  For the latter two the
correlation distributions are simultaneously projected onto
{\em difference variables}
$\phi_{\Delta}$ and $\eta_{\Delta}$; the projection is then
referred to as a {\em joint autocorrelation}.

\section{Charge-Independent Joint Autocorrelations on $(\eta_\Delta,\phi_\Delta)$}\label{axialCI62}
 
\begin{figure}[htb]
                 \insertplotx{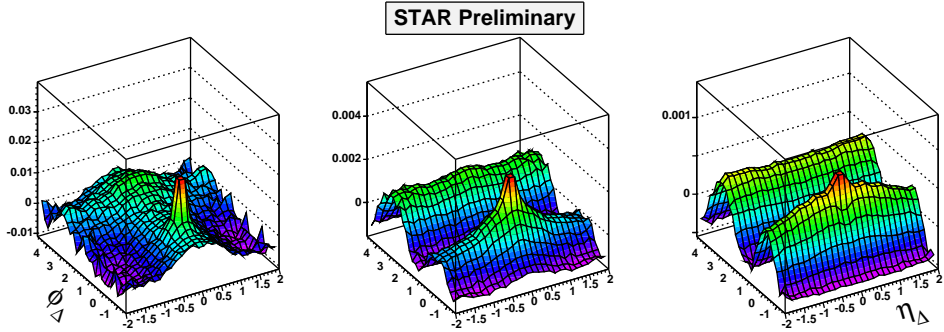}
\vspace*{-1.0cm}
\caption[]{
$\Delta\rho/\sqrt{\rho_{ref}}$ for Au-Au collisions at $\sqrt{s_{NN}}$ = 62 GeV
on $(\eta_\Delta,\phi_\Delta)$
for centralities $\sim$80-90\%, 50-60\% and 10-20\% fraction of Au-Au total
cross section.}
\label{fig1}
\end{figure}

Plotted in Fig.~\ref{fig1} are perspective views of preliminary
charge-independent
joint autocorrelations $\Delta\rho/\sqrt{\rho_{mix}}$ on difference variables
$(\eta_\Delta,\phi_\Delta)$
for three centrality bins (approximately 80-90\%, 50-60\% and 10-20\%)
for Au-Au collisions
at 62 GeV.
The distributions are dominated by 1) a $\cos(2\phi_\Delta)$ component
attributed to elliptic flow;
2) a $\cos(\phi_\Delta)$ component associated with transverse
momentum conservation, and 3) a 2D same-side
($|\phi_\Delta| < \pi / 2$) peak, which is the principal object of interest
here, and
assumed to be associated with low-$Q^2$ scattered parton fragmentation.
In addition, for the most-peripheral data, a $\phi_\Delta$ independent gaussian
distribution on $\eta_\Delta$ is observed similar to that seen in
correlation studies of p-p collision data~\cite{jeffpp} and assumed to be due
to charge-ordering associated with longitudinal string fragmentation.
This feature vanishes quickly with centrality.
The same-side peak in Fig.~\ref{fig1} varies strongly with centrality,
transitioning from nearly symmetric on $(\eta_\Delta,\phi_\Delta)$ for
peripheral collisions to dramatically broadened along $\eta_\Delta$ and
narrowed on $\phi_\Delta$ for the
more central collisions.
Resonance ({\em e.g.} $\rho^0 , \omega$) decays contribute about 3\% of the 
peaks at (0,0)~\cite{axialCI}.  Electron conversion pairs
which remain after the $dE/dx$ cuts add to
the bin at (0,0) producing the narrow spike there.

\begin{figure}[htb]
                 \insertplotxx{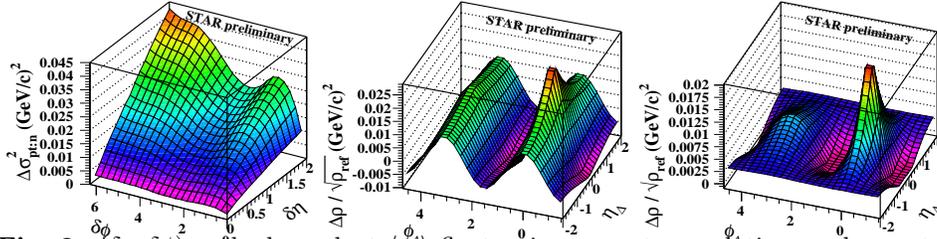}
\vspace*{-1.0cm}
\caption[]{$(\delta \eta,\delta \phi)$ scale-dependent
$\langle p_t \rangle$ fluctuation, $p_t$ autocorrelation, and $p_t$
autocorrelation with $\cos(2 \phi_{\Delta})$ and $\cos(\phi_{\Delta})$
terms subtracted for mid-central
(20-30\%) Au-Au collisions at 200 GeV. $\rho_{ref}$ is the same as
$\rho_{mix}$ in the text.
}
\label{fig2}
\end{figure}

Mean-$p_t$ fluctuation measure $\Delta \sigma^2_{p_t:n}(\delta \eta,\delta \phi)$~\cite{meanptprc}
obtained for particles within a two-dimensional bin of size ($\delta \eta,\delta \phi$),
is equivalent to an
integral of $p_t$ autocorrelation
$\Delta \rho(p_t:n;\eta_\Delta,\phi_\Delta)  / \sqrt{\rho_{mix}}$
over difference variables from (0,0) to $(\delta \eta,\delta \phi)$~\cite{ptscale}.
Inversion of this integral equation yields
$p_t$ autocorrelations. The latter
provide access to temperature and/or velocity distributions
in the colliding system independent of that afforded by number of pair
correlations~\cite{ptscale}.  Fig.~\ref{fig2} shows
$\Delta \sigma^2_{p_t:n}(\delta \eta,\delta \phi)$
(left panel), the
resulting $p_t$ autocorrelation (middle panel), and $p_t$ autocorrelation
with $\cos(2 \phi_{\Delta})$ and $\cos(\phi_{\Delta})$ subtracted~\cite{ptscale}
(right panel).  The same-side peak structure broadens dramatically with
centrality (not shown) on $\eta_{\Delta}$ as in Fig.~\ref{fig1} but remains
narrower than the corresponding number of pair correlations.

\section{Conclusions}
In STAR we are studying
how the same-side autocorrelation structures
evolve with centrality in Au-Au collisions at $\sqrt{s_{NN}}$ = 62, 130
and 200 GeV. 
Same-side
charge-average pair number and $p_t$ autocorrelations strongly elongate on
$\eta$ and narrow on $\phi$.
The STAR experiment results reported here and
elsewhere~\cite{ptscale,mtxmtCI,axialCI}
are consistent with a picture in which (1) semi-hard parton
scattering transfers momentum ($p_t$) to the bulk medium, inducing
temperature/velocity fluctuations in the soft $p_t$ range
whose correlation amplitudes
scale with the number of binary collisions per participant, and
(2) particles correlated with semi-hard scattered partons in the early stages of
the collision interact with the longitudinally expanding medium 
resulting in larger average $|\eta_{\Delta}|$.
The reduced width on $\phi$ is not understood.


\vfill\eject
\end{document}